\documentclass[letter]{aa} 
\usepackage{booktabs,caption}
\usepackage[flushleft]{threeparttable}
\usepackage{graphicx}
\usepackage{txfonts}
\usepackage{color}

\begin{document}

   \title{Gaia DR2 reveals a star formation burst in the disc 2-3 Gyr ago}

 \author{
R.\,Mor\inst{1} 
\and A.C.\, Robin\inst{2}
\and F.\, Figueras \inst{1}
\and S.\, Roca-F\`abrega \inst{3}
\and X. Luri \inst{1}  
          }

 \institute{
Dept. F\'isica Qu\`antica i Astrof\'isica, Institut de Ci\`encies del Cosmos, Universitat de Barcelona (IEEC-UB), Mart\'i Franqu\`es 1, E08028 Barcelona, Spain.
\email{rmor@fqa.ub.edu}
\and
Institut Utinam, CNRS UMR6213, Universit\'e de Bourgogne Franche-Comt\'e, OSU THETA , Observatoire de Besan\c{c}on, BP 1615, 25010 Besan\c{c}on Cedex, France
\and
Departamento de F\'isica de la Tierra y Astrof\'isica, Facultad de Ciencias F\'isicas, Plaza Ciencias, 1, Madrid, E-28040, Spain
}

   \date{Received Month dd, yyyy; accepted Month dd, yyyy}

 
  \abstract
 {We use Gaia data-release 2 (DR2) magnitudes, colours, and parallaxes for stars with G<12 to explore a parameter space with 15 dimensions that 
simultaneously includes the initial mass function (IMF) and a non-parametric star formation history (SFH) for the Galactic disc. This inference is performed by combining the Besan\c{c}on Galaxy Model fast approximate simulations (BGM FASt) and an approximate Bayesian computation algorithm. We find in Gaia DR2 data an imprint of a star formation burst 2-3 Gyr ago in the Galactic thin disc domain, and a present star formation rate (SFR) of $\approx1M_\odot/yr$. Our results show a decreasing trend of the SFR from 9-10 Gyr to 6-7 Gyr ago. This is consistent with the cosmological star formation quenching observed at redshifts $z<1.8$. This decreasing trend is followed by a SFR enhancement starting at $\sim 5 Gyr$ ago and continuing until $\sim 1 Gyr$ ago which is detected with high statistical significance by discarding the null hypothesis of an exponential SFH with a p-value=0.002. We estimate, from our best fit model, that about $50\%$ of the mass used to generate stars, along the thin disc life, was expended in the period from 5 to 1 Gyr ago. The timescale and the amount of stellar mass generated during the SFR enhancement event lead us to hypothesise that its origin, currently under investigation, is not intrinsic to the disc. Thus, an external perturbation is needed for its explanation. Additionally,  for the thin disc we find a slope of the IMF of $\alpha_3 \approx 2$ for masses   $M>1.53M_\odot$ and $\alpha_2 \approx 1.3$ for the mass range  between 0.5 and 1.53 $M_\odot$. This is the first time that we consider a non-parametric SFH for the thin disc in the Besan\c{c}on Galaxy Model. This new step, together with the capabilities of the Gaia DR2 parallaxes to break degeneracies between different stellar populations, allow us to better constrain the SFH and the IMF.}


   \maketitle
%
 
\section{Introduction}\label{Sec1}

The star formation history (SFH) of the Milky Way disc contains essential  information to understand the Galactic structure and evolution, including key information of its merger history (e.g. \citealt{Gilmore2001}). Recently, \cite{Antoja2018} discovered, using Gaia data, that an external interaction perturbed the Galactic disc in the last billion years. Moreover, \cite{Helmi2018} suggested that a merger led to the formation of the thick disc. Furthermore, from the cosmological simulations in the framework of $\Lambda CDM$, it is known that the probability that a Milky Way-like Galaxy had a minor merger in the last 10 Gyr is high (e.g. \citealt{Stewart2008}). These mergers can trigger stellar formation that we expect to detect in the observational catalogues when characterising the SFH of the Galactic disc (e.g. \citealt{Kruijssen2018}). The analysis of the Milky Way SFH cannot be disentangled from the study of the stellar initial mass function (IMF), as discussed in \cite{Haywood1997} and \cite{Aumer&Binney2009}, for example. In this context, the unprecedented accuracy of the Gaia data-release 2 (DR2) data \citep{GaiaPrusti,GaiaDR2} represents a great opportunity to search for hints of star formation bursts in the Galaxy using the population synthesis Besan\c{c}on Galaxy Model (BGM; \citealt{Robin2003}). Previous studies performed with BGM used catalogues of colours and apparent magnitudes to perform parameter inference (e.g. \citealt{Robin2014}, \citealt{Mor2018}), carrying some degeneracies mostly due to the lack of information of the intrinsic luminosity of the stars. Now, for the first time, Gaia parallaxes help us to break some of the degeneracies between different stellar populations for a large stellar sample. Following the approach proposed in \cite{Mor2018} here we compare synthetic versus observed full-sky magnitude-limited stellar samples by using a Bayesian approach to simultaneously explore a non-parametric SFH, a three truncated power-law IMF, and the disc density laws. This is  the first time that, using BGM, the SFH of the thin disc is considered non-parametric. In practice this means that we infer the surface star formation rate (SFR) of nine age bins from 0 to 10 Gyr.
We summarise our method in Sect. \ref{Sec2}. In Sect. \ref{Sec3} we present the observational sample used and in Sect. \ref{Sec4} we discuss the analysis of the data. The resulting SFH and IMF are presented and discussed in Sect. \ref{Sec5}. Finally, in Sect. \ref{Sec6} we present the conclusion.  

\section{BGM FASt for Gaia}\label{Sec2}

  We use an approximate Bayesian computation (ABC) algorithm \citep{Jennings2017} together with the Besan\c{c}on Galaxy model fast approximate simulations (BGM FASt, \citealt{Mor2018}) to infer a parameter space with 15 dimensions. BGM FASt is an analytical framework to perform very fast Milky Way simulations based on BGM. The theory of BGM FASt and the basis of the parameter inference strategy that we use in this work is extensively described in \cite{Mor2018}. Summarising, our iterative parameter inference strategy works as follows. First we sample a set of 15 parameters from the prior probability distribution functions (PDFs); we choose these to be  wide Gaussians centred on the results of \cite{Mor2018} (see their Figs. 6 and 7). Subsequently, we perform a new BGM FASt simulation using the sampled parameters as inputs. We then define $M_\varpi$ as a combination of Gaia observables: $M_\varpi= G + 5\cdot log_{10}(\varpi/1000)+5$,  where $\varpi$ is the parallax of the star. If the parallax accuracy and the interstellar absorption go to zero ($\sigma_\varpi \to 0$ and $A_G \to 0$),  the $M_\varpi$ becomes the absolute magnitude of the star. We then use the Poissonian distance metric\footnote{We know from \cite{Kendall1973} and \cite{Bienayme1987} that the Poissonian distance is a good choice for the comparison. For simplicity it can be understood as a goodness-of-fit: the shorter the Poissonian distance, the closer 
the simulation to the observations.} ($\delta_P$; Eq. 58 from \citealt{Mor2018}) to compare synthetic versus Gaia DR2 $M_\varpi$-colour ($G_{Bp}-G_{Rp}$) distributions for the whole sky divided into three latitude ranges ($|b|<10$, $10<|b|<30$ and $30<|b|<90$). If the resulting Poissonian distance is smaller than an imposed threshold, the given set of 15 parameters is accepted as part of the posterior PDF. Otherwise it is rejected. We set the threshold to be small enough to ensure that we discard all the combinations of parameters that give worse results than the best model in  \cite{Mor2018}.

\begin{table}
\caption{Age intervals, priors, and posterior PDFs (see Sects. \ref{Sec4} and \ref{Sec5}) for the 15 inferred parameters for our fiducial case (see text). }
\label{Table1}
\centering
\resizebox{\columnwidth}{!}{
\begin{tabular}{c c c c c c}        
\hline\hline                 
parameter &   units& age (Gyr)  & $\mu_{S}$ & $\sigma_S$ &  posterior\\    
\hline  
\vspace{1.5pt}
   $\Sigma^1_\odot$ &$M_\odot pc^{-2}$ &0-0.1& 0.17 &0.5  & $0.16^{+0.07}_{-0.04}$ \\
   \vspace{1.5pt}
$\Sigma^2_\odot$ &$M_\odot pc^{-2}$ &0.1-1 &1.62 & 3.0   & $2.2^{+0.4}_{-0.5}$\\
   \vspace{1.5pt}
  $\Sigma^3_\odot$&$M_\odot pc^{-2}$ &1-2& 2.07 &3.0&$6.5^{+1.4}_{-1.4}$\\
   \vspace{1.5pt}
  $\Sigma^4_\odot$ &$M_\odot pc^{-2}$&2-3 & 2.39 &3.0 &$8.7^{+2.7}_{-2.1}$\\
   \vspace{1.5pt}
  $\Sigma^5_\odot$ &$M_\odot pc^{-2}$ & 3-5 &5.92 &6.0&$12.1^{+4.1}_{-4.5}$\\
   \vspace{1.5pt}
  $\Sigma^6_\odot$ &$M_\odot pc^{-2}$ &5-7 &7.86 &8.0 &$7.7^{+5.7}_{-4.1}$\\    
     \vspace{1.5pt}
  $\Sigma^7_\odot$&$M_\odot pc^{-2}$ &7-8& 5.62&6.0 &$5.8^{+7.2}_{-2.9}$\\
   \vspace{1.5pt}
  $\Sigma^8_\odot$&$M_\odot pc^{-2}$&8-9 & 5.62&6.0&$7.4^{+5.7}_{-4.4}$\\
   \vspace{1.5pt}
  $\Sigma^9_\odot$&$M_\odot pc^{-2}$&9-10&5.62 &6.0&$9.8^{+5.9}_{-5.3}$\\
   \vspace{1.5pt}
   $\rho^{young}_\odot \cdot 10^{-3}$ &$M_\odot pc^{-3}$&$\approx 10$& $3.6$  &$3.6$ &$2.6_{-0.4}^{+0.7}$\\
   \vspace{1.5pt}
   $\rho^{old}_\odot\cdot 10^{-3}$&$M_\odot pc^{-3}$ &$\approx 12$& $0.5$&0.5 &$0.6_{-0.3}^{+0.7}$\\
   \vspace{1.5pt}
   $\alpha_1$ &-&all &0.5 &0.5   &$-0.5^{+0.8}_{-0.5}$\\
   \vspace{1.5pt}
   $\alpha_2$ &-&all& 2.1&0.5     &$1.3^{+0.3}_ {-0.3}$ \\
   \vspace{1.5pt}
   $\alpha_3$ &- &all & 2.9&0.5 &$1.9^{+0.2}_{-0.1} $\\
   \vspace{1.5pt}
   $h_R$ & pc & 0.10-10& 2151&274 & $1943 ^{+190}_{-370}$\\ 
        \\ 
\hline                                   
\end{tabular}%
}
\begin{tablenotes}
      \small
      \item \textbf{Notes.} The prior PDFs are Gaussians centred on $\mu_{S}$ with variance $\sigma^2_S$. These PDFs are truncated at 0 except for the slopes of the IMF. The $\mu_S$ are taken from Fig. 7 of \cite{Mor2018}. The $\mu_S$ of the nine $\Sigma^j_\odot$ are obtained by integrating, for each age interval, the exponential SFH given in the mentioned figure. For the eleven density parameters the $\sigma_S$ is chosen big enough to allow the 0 to be inside 1$\sigma$. The $\sigma_S$ for the $h_R$ is chosen to be the same as resulting in \cite{Mor2018}. For the three slopes of the IMF the $\sigma_S$ is set to $0.5$.
    \end{tablenotes}
\end{table}

We infer the 15 parameters listed in Table \ref{Table1} which are the following: the thin disc radial scale length ($h_R$) for populations older than 0.10 Gyr; the three slopes of a three truncated power-law IMF, $\alpha_1$ (for $0.09M_\odot< M <0.5M_\odot$), $\alpha_2$ (for $0.5M_\odot< M<1.53M_\odot$), and $\alpha_3$ (for $1.53M_\odot<M<120M_\odot$); the present volume stellar mass density of the thick disc at the position of the Sun for the BGM young ($\rho^{young}_\odot$)  and old ($\rho^{old}_\odot$) components of the thick disc \citep{Robin2014}; and the surface stellar mass density at the position of the Sun ($\Sigma^j_\odot$) of the generated stars along the life of the thin disc, for nine intervals of age. These nine values of $\Sigma^j_\odot$ divided by the interval of age become the mean surface SFR per age bin ($M_\odot Gyr^{-1} pc^{-2}$). All of them together constitute a non-parametric SFH. Each complete and robust inference of the full set of parameters requires $2\cdot 10^4$ CPU hours in the Spark environment of the Big Data platform at the University of Barcelona. We used more than $10^5$ hours of CPU.

The fixed model ingredients are described in \cite{Mor2018} following \cite{Robin2003}, \cite{Robin2012}, and \cite{Czekaj2014}. We adopt the photometric transformation of \cite{Evans2018} to transform the simulated data from Johnson to Gaia bands. The error modelling of astrometric and photometric data and the angular resolution of the stellar multiple systems (0.04 arcsec) are chosen accordingly to \cite{GaiaDR2}. We define as our fiducial case the one that uses a non-parametric SFH and the Stilism extinction map \citep{Rosine2018}. This is the most recent extinction map specifically developed to be used in BGM. The prior PDFs adopted for our fiducial case are shown in Table \ref{Table1}.

\section{Gaia DR2 observational sample}\label{Sec3}

We use, from Gaia DR2, the G mean magnitude, the colours $(G_{Bp}-G_{Rp}),$ and the parallaxes ($\varpi$) for a full-sky sample limited to stars with magnitude G<12. The completeness of the sample is estimated using the pre-cross-match of Gaia DR2 with Tycho-2 catalogue from the Gaia archive. First, we take all stars in this cross-match with $V_T<11$, where Tycho-2 is $99\%$ complete \citep{Hog2000}. We then compare the obtained number of stars with the number of stars in Tycho-2 with the same magnitude limit. The results obtained show that the cross-match has $\sim 2\%$ less stars than Tycho-2. Additionally, Gaia DR2 is known to be complete from G=12 to G=17 \citep{GaiaDR2}. Therefore, it is plausible to assume that, for G<12, the catalogue is more complete closer to G=12. As a consequence, we expect that for $G<12$ the catalogue is at least as complete as for $V_T<11$. Therefore, we estimate that the Gaia DR2 catalogue is about $97\%$ complete up to $G= 12$. Additionally, we feel it necessary to mention that about $1\%$ of the data have either no colours or have no parallax. We also limit our model-versus-data comparison in the colour range where the photometric transformation of \cite{Evans2018} is valid; this is the range of colour ($G_{Bp}-G_{Rp}$) from $-0.47$ to $2.73$. To avoid the white and brown dwarfs, which for the moment are treated independently of the thin disc in BGM, we consider only stars with $M_\varpi<10$. The total number of stars in the Gaia DR2 subsample used is 2890208.

\begin{table}
\caption{Summary of the considered model variants, the adopted SFH, extinction map, data used for the fitting and Poissonian distance ($\delta_P$).} 
\label{Table2}      %
\centering  
\resizebox{\columnwidth}{!}{%
\begin{tabular}{c c c c c}        
\hline\hline                 
model & SFH & extinction  & fitted to & $\delta_P$  \\    
\hline   
 MP-S  & exponential  & Stilism &Tycho-2 V<11 & $7.5 \cdot 10^5$ \\
 G12Exp-S& exponential & Stilism & Gaia DR2 G<12 & $7.4 \cdot 10^5$ \\
\textbf{G12NP-S} & \textbf{non-param.} & \textbf{Stilism} &  \textbf{Gaia DR2  G<12} &  $\boldsymbol{5.6 \cdot 10^5} $\\
   G12NP-D &  non-param. &  Drimmel & Gaia DR2 G<12& $5.4 \cdot 10^5$ \\

  G12NP-M  & non-param. & Marshall &Gaia DR2 G<12& $6.0 \cdot 10^5$ \\

        \\ 
\hline                                   
\end{tabular}%
}
\begin{tablenotes}
      \small
      \item \textbf{Notes}. The G12NP-S is our fiducial case. The MP-S model was derived in \cite{Mor2018} (see their Fig. 7), its slopes of the IMF are $\alpha_1=0.5$, $\alpha_2=2.1$ and $\alpha_3=2.9$. The extinction maps named Stilism, Drimmel, and Marshall are from \cite{Rosine2018}, \cite{Drimmel2001}, and \cite{Marshall2006}, respectively.
    \end{tablenotes}
\end{table}

\section{Analysis of the data}\label{Sec4}

In Table \ref{Table2} we present the model variants considered here and the resulting Poissonian distance for each one. The MP-S was obtained from a fit to Tycho-2 photometry and its main parameters are reported in Fig. 7 of \cite{Mor2018}. The remaining model variants in the present work  result from fitting Gaia DR2 data using photometric data and parallaxes. As shown in this table, the Poissonian distance of the models that use a non-parametric SFH is smaller than that of the models with an exponential SFH. We want to emphasise the fact that the difference in the Poissonian distance between G12NP-S (our fiducial case; see Sect. \ref{Sec2}) and G12NP-D is not large enough to settle on which extinction map is better.  

\begin{figure*}
   \centering
 
   \includegraphics[width=\hsize]{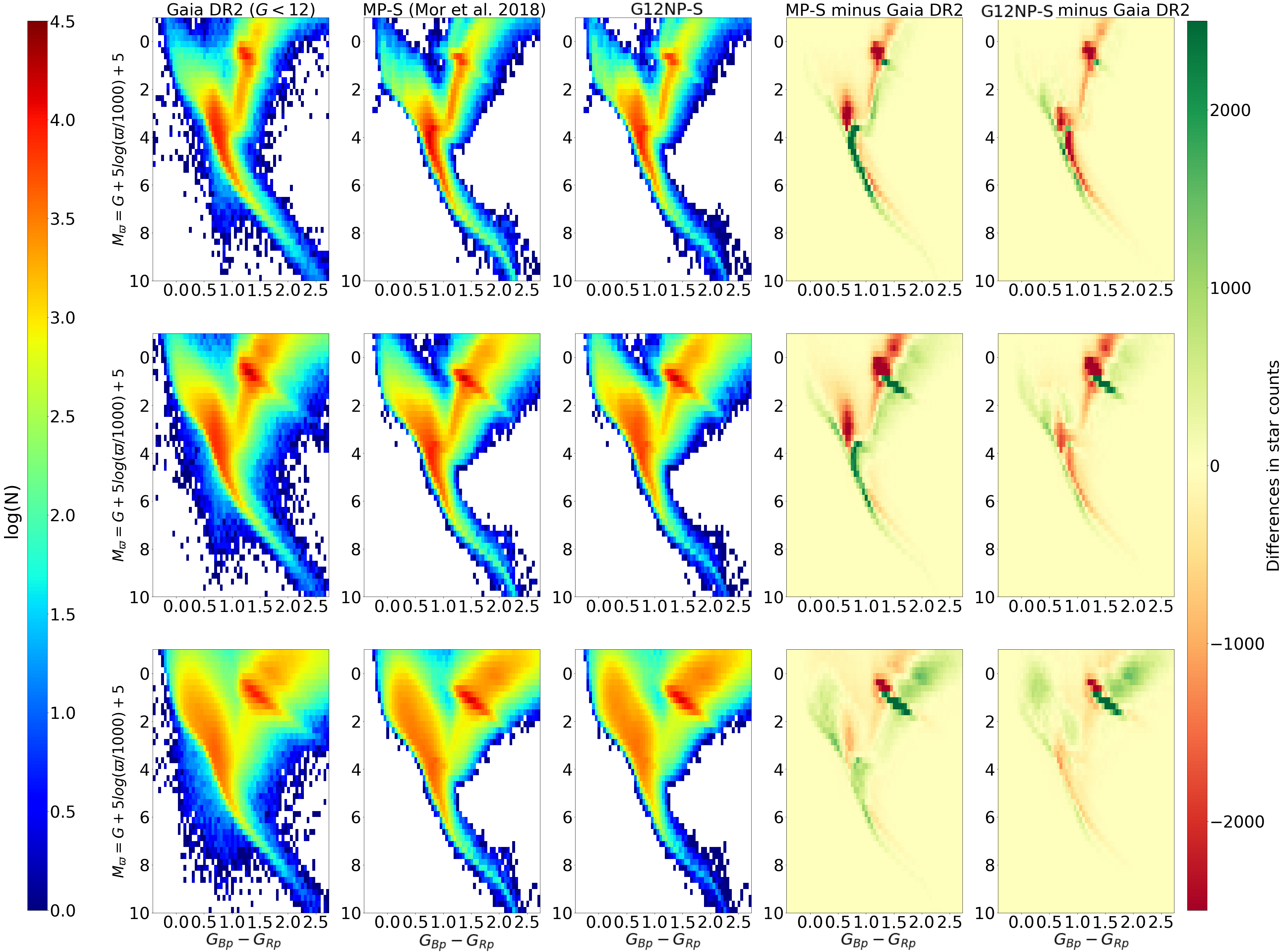}
      \caption{ $M_\varpi$ vs. Gaia colour $G_{Bp}-G_{Rp}$ for the stars with G<12 divided into three latitude ranges:  \textbf{first row:} $30<|b|<90$; \textbf{second row:} $10<|b|<30$; \textbf{third row} $|b|<10$. The colour-map of the first, second, and third columns shows the logarithm of the star counts in each bin. The \textbf{first column} is Gaia DR2  data and the  \textbf{second column} is the most probable model variant from \cite{Mor2018}, which has an exponential SFH and whose IMF has $\alpha_3 \approx 3$ .The third is for the best-fit model using a non-parametric SFH, whose IMF has $\alpha_3 \approx 2$ (this work). The BGM simulations performed for this figure use the Stilism extinction map. In the \textbf{fourth column} we show, for each bin, the difference of star counts MP-S minus Gaia DR2 data. In the \textbf{fifth column} we show, for each bin, the difference of star counts G12NP-S minus Gaia DR2 data. Observational data and simulations are limited here at $G < 12$ and $10<M_\varpi <-1$.}
         \label{CMDStilism}
   \end{figure*}

In Fig. \ref{CMDStilism} we show a density map, $M_\varpi,$ as a function of the Gaia colour $(G_{Bp}-G_{Rp})$ for the three latitude ranges considered. For the stars with $10<M_\varpi<-1$ we set the bin size to 0.05 mag in colour and 0.25 mag in $M_\varpi$. For stars with $M_\varpi<-1$ the bin size is enlarged to 0.5 mag in colour and 1 mag in $M_\varpi$ to allow a robust statistical analysis; these stars, which represent $7\%$ of the sample, are not shown in the figure. Even  though $M_\varpi$ is not strictly the absolute magnitude, we refer to this density map as if it were a true Hertzsprung-Russel diagram. The first column shows the Gaia DR2 data, the second column shows the MP-S model variant, and the third column shows the best-fit model obtained in this work for our fiducial case (G12NP-S). The fourth and fifth columns show the differences, in star counts per bin, between our Gaia DR2 sub-sample and both the old MP-S and the new G12NP-S. With this new fit we improve the agreement of the model with the data for the three latitude ranges. Focusing on the bright end ($M_\varpi<-1$), the improvement is significant mostly in the Galactic plane.

The first feature of Fig. \ref{CMDStilism} that we want to comment on is the following. In the Gaia DR2 data, we can see a blob of stars (mostly in blue) below the main sequence (MS) that is not reproduced in the simulations. These are stars that are flagged with a bad colour-excess \citep{Evans2018} in the Gaia catalogue and represent only 0.1\% of our sample. By comparing the fourth and fifth columns of Fig. \ref{CMDStilism}, we can see how the agreement with Gaia data is better when using the non-parametric SFH (G12NP-S). Both the excess of stars detected in the MP-S around the MS region and the deficit of stars in the region with $0.5<(G_{Bp}-G_{Rp})<1.0$ and $1.5<M_\varpi<3 $ are clearly diminished in the new G12NP-S.
 
The high quality of the new information given by Gaia data reveals new discrepancies between model and data. Most of these discrepancies could come from several assumptions on the fixed ingredients of the BGM model that, as largely discussed in Sect. 7.3 of \cite{Mor2018}, can impact our parameter inference. Here we discuss the discrepancies that we see in the fifth column of Fig. \ref{CMDStilism}, grouped into three main areas: the red giant branch (RGB), the MS, and the region of stars with $3<M_\varpi<4$ and $0.5<(G_{Bp}-G_{Rp})<1,$ hereafter referred to as the square region. In the RGB region, we notice that in the simulations the position of the RGB clump is shifted and that it is less extended than in the observations. In this region, the asymptotic giant branch bump is less dense in the simulations; this effect is mostly seen at intermediate latitudes. In the MS region, we detect a clear sequence where the simulation has an excess of stars and, immediately above, a clear sequence with a deficit of stars. We also notice that in the square region the simulation has a deficit of stars. This last effect is stronger at high and intermediate latitudes. As expected, these differences are caused by a mixture of factors. From a first analysis of our simulations we conclude that the thick disc modelling and the stellar evolutionary models are the two main ingredients causing the discrepancies. There are other ingredients that can contribute to these discrepancies however: the assumed colour transformation; the radial metallicity distribution; the age-metallicity relation; and the atmosphere models, which mostly have their impact in the MS and RGB; the rate of mass loss assumed in the stellar evolutionary models; the assumption of the extinction map, mostly affecting the RGB; and finally, the assumed resolution of the stellar multiple systems mostly affecting the MS. Work is in progress to more thoroughly analyse these discrepancies, to confirm their causes, and to improve the BGM model accordingly.

\section{The resulting SFH and IMF}\label{Sec5}

\begin{figure}
   \centering

   \includegraphics[width=\hsize]{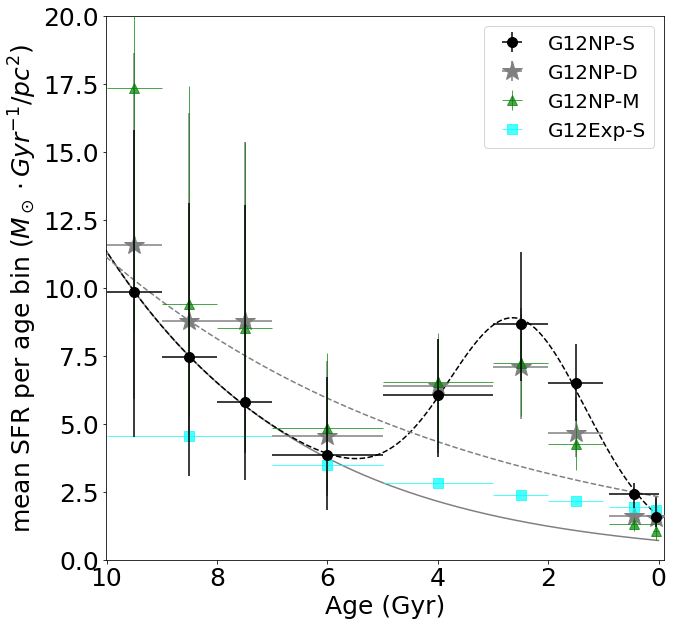}
      \caption{Most probable values of the mean SFR for the age bin obtained from the posterior PDF. The vertical error bars indicate the 0.16 and 0.84 quantiles of the posterior PDF. The horizontal error bars indicate the size of the age bin. The grey and black dashed lines are, respectively, an exponential function and a distribution formed by a bounded exponential plus a Gaussian, fitted to the G12NP-S results. The grey solid line is the exponential part of this exponential plus Gaussian fit. See Table \ref{Table2} for details of the SFH and extinction maps used.}
         \label{sfhhint}
   \end{figure}

In the last column of Table \ref{Table1} we show the results of the 15 inferred parameters for our fiducial case. In this section we focus on the discussion of the resulting IMF \footnote{As largely discussed in \cite{Mor2017}, the IMF considered in BGM is a composite
IMF (or Integrated Galactic IMF; IGIMF). } and the non-parametric SFH of the thin disc. In Fig. \ref{sfhhint} we present the nine values of the local mean surface SFR as a function of age that constitute the SFH. In this figure we show the results for our fiducial case (G12NP-S). Additionally, to evaluate the impact of the choice of the extinction model in our results, we present the inferred SFH when using the  \cite{Drimmel2001} (G12NP-D) and \cite{Marshall2006} \footnote{The \cite{Marshall2006} extinction map covers the longitude ranges $-100<l<100$ and the latitude ranges $|b|<10$. Therefore, Drimmel map is used for the rest of the sky.} (G12NP-M) extinction maps (see Table \ref{Table2}). We notice that the differences in the SFR among them are not larger than $1.5\sigma$. Regardless of the choice of the extinction map we can see a general decreasing trend from 9-10 Gyr to 6-7 Gyr ago followed by a SFR enhancement event beginning about 5 Gyr ago. This enhancement event is of about 4 Gyr in duration with a maximum at about 2-3 Gyr ago and with a final decreasing trend until the present time. We would like to point out that our results do not rule out an earlier beginning for this SFR enhancement event, nor a constant (or slightly decreasing) SFR with a value of about $7M_\odot \cdot Gyr^{-1} pc^{-2}$ from 10 Gyr ago until 1 Gyr ago, with a very sharp and fast drop in the last 1 Gyr. We estimate from our best-fit model that about $50\%$ of the mass used to generate stars throughout the life of the thin disc was expended in the period from 5 to 1 Gyr ago. 

To evaluate the statistical significance of the enhancement event we compare the values of the non-parametric SFH of our fiducial case (G12NP-S)  with both: (1) the values of the SFH obtained when imposing an exponential SFH in BGM and performing the fit with Gaia data (G12Exp-S results) and (2) the values of an exponential shape fitted to the G12NP-S results using the least squares method (grey dashed line in Fig. \ref{sfhhint}). This fitted exponential shape is purely mathematical and does not necessary make physical sense. The statistical significance of the points in the SFR enhancement event for the first and second cases are the following:  $2.8\sigma$  and $2.5 \sigma$ for the point at 2.5 Gyr ago (the relative maximum);  $ 3\sigma$ and $1.5\sigma$ for the point at 1.5 Gyr ago, and finally $1.5\sigma$ and $0.8\sigma$ for the point at 4 Gyr ago. To evaluate the significance of the event as a whole we perform two tests with the following two null hypotheses: (1) the SFH is exponential and follows the results of the G12Exp-S variant and (2) the SFH follows the fitted mathematical exponential shape (grey dashed line in Fig. \ref{sfhhint}). In a subsequent step, for both tests, we assume that the distribution of the SFR at each age bin follows a Gaussian centred in the values given by the null hypothesis and with $\sigma$ estimated from the obtained posterior PDF. Afterwards, for both null hypotheses, we compute the p-value for the points at 1.5, 2.5, and 4 Gyr ago. We finally obtain a p-value of the global event by combining the individual p-values using Fisher's Method. The results give p-values of <0.001 and =0.002, for hypotheses (1) and (2), respectively; we therefore reject both null hypotheses. 

To mathematically characterise the SFR enhancement event we fit a bounded exponential plus a Gaussian function to 
the results (black dashed line in Fig. \ref{sfhhint}), obtaining for the Gaussian component $\mu=2.57 Gyr$ and $\sigma=1.25 Gyr$. In Fig. \ref{sfhhint} we additionally show the exponential part of this last fit (grey solid line) where we see how the SFH of our fiducial case follows an exponential shape from 10 Gyr until 6-7 Gyr ago. From all the performed tests we conclude that the SFR enhancement that we find is statistically significant. 
  
 Our findings that the thin disc SFH does not follow a simple decreasing shape until the present are in good agreement with \cite{Snaith2015}, \cite{HaywoodMateo2016}, and \cite{Haywood2018} who found, using data with metallicities and assuming a fixed IMF, the existence of an SFR quenching followed by a reactivation. \cite{Kroupaclusters},  using stellar kinematics,  found the SFH to behave similarly. The relative maximum of the SFR that we find at 2-3 Gyr ago is compatible with the results of \cite{Vergely2002} and \cite{Cignoni2006} that, using Hipparcos data in a sphere of 80 pc around the Sun and assuming a fixed IMF, found maximum peaks at 1.75-2 Gyr ago and 2-3 Gyr ago, respectively. Recently, \cite{Bernard2018}, in a tentative work using TGAS data, pointed towards the existence of a relative maximum also located 2-3 Gyr ago.  
 
\begin{figure}
   \centering

   \includegraphics[width=\hsize]{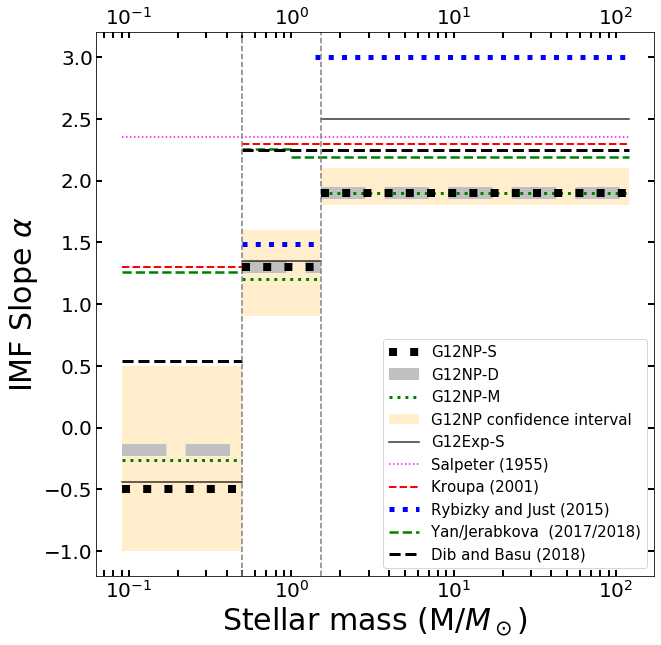}
      \caption{ Values of the slopes of the IMF obtained in this work together with a compilation of results in the literature. The dotted vertical lines indicate the mass limits of the three truncated power-law IMF that we adopt here ($x_1=0.5$ and $x_2=1.53$). See Table \ref{Table2} for details on the SFH and extinction maps used.}
         \label{IMF}
   \end{figure}

 In Fig. \ref{IMF} we show the resulting slopes $(\alpha)$ of the inferred IMF as a function of stellar mass and a compilation of results in the literature. For the mass range between $0.5M_\odot$ and $1.53M_\odot$ we find $\alpha_2=1.3\pm 0.3$, in very good agreement with \cite{Rybizki2015} who found $\alpha=1.49\pm 0.08$ (in the range $0.5M_\odot<M<1.4M_\odot$). For masses larger than $1.53M_\odot$ we find $\alpha_3=1.9_{-0.1}^{+0.2}$, which is flatter than the $\alpha_3$ obtained by \cite{Salpeter1955} and \cite{KroupaSci2002}. For the low-mass range ($0.09M_\odot <M<0.5M_\odot$) we obtain values between $\alpha_1=-1$ and $\alpha_1=0.5$. We must keep in mind our estimation that about  $99.6\%$ of the stars in our sample have masses between $0.5M_\odot$ and $10M_\odot$, with only $0.1\%$ of the stars in our sample belonging to the lowest mass range, and that of the order of $10^4$ stars have $M>10M_\odot$. We also want to compare our results with two works that consider a non-universal IMF. These are the recent works of \cite{Dib2018} and \cite{Jerabkova2018} (see Appendix \ref{AppIMF}). We note that, as in our results, their IMFs have a shallower shape than the  values of Kroupa or Salpeter.
 The information from Gaia parallaxes, when imposing an exponential SFH, brings the resulting $\alpha_3$ to be $\approx 2.5$. This is flatter than in our previous works \citep{Mor2017,Mor2018} but compatible with the $\alpha_3 \approx 2.7$ of the IGIMF (e.g. \citealt{Kroupa2013}). When we adopt a non-parametric SFH we find $\alpha_3 \approx 2$, more in the direction of \cite{Zonoozi2019}. We know from \cite{Mor2018} (e.g. their Fig. 7) that when characterising the IGIMF from star counts the correlation of $\alpha_3$ with the SFH is high. In our previous works, the fact that we were imposing an exponential SFH resulted in a steeper $\alpha_3$. These correlations between the $\alpha_3$ and the SFH are also observed in our present work. We find that the $\alpha_3$ is anti-correlated with the four $\Sigma^j_\odot$ values for the age bins from 0.1 to 5 Gyr, with a Pearson's correlation coefficient going from about $-0.8$ to about $-0.5$. We also note that these four surface densities ($\Sigma^j_\odot$) are also correlated among them, with coefficients from about 0.3 to 0.5. We want to emphasise that the effects of these correlations in the results (Table \ref{Table1}, Figs. \ref{sfhhint} and \ref{IMF}) are already taken into account when we provide the 0.16 and 0.84 quantiles of the posterior PDFs. From the surface SFR of the youngest population ($1.6_{-0.4}^{+0.7} M_\odot Gyr^{-1} pc^{-2}$) we find the present SFR in the disc to be about $1M_\odot/yr$. This result is very sensitive to the disc scaling of the youngest population. We also find a radial scale length of $h_R=1943^{+190}_{-370}$ compatible with \cite{Robin2012}. For robustness we repeated our analysis by adding to the Gaia DR2 parallaxes the offset of 0.029 mas reported in \cite{Lindegren2018}, concluding that the impact on the derived IMF and SFH is much smaller than the impact of the choice of the extinction map.

\section{Conclusion}\label{Sec6}

For the first time we have considered a non-parametric SFH for the thin disc in the BGM model. This new step, together with the capability of the Gaia DR2 parallaxes to break degeneracies between different stellar populations, allowed us to better constrain the thin disc SFH and IMF. 
The resulting SFH shows a decreasing trend from 9-10 to 6-7 Gyr ago that is consistent with the quenching observed in a cosmological context for redshifts $z<1.8$ (e.g. \citealt{Rowan2016}) and is also compatible with the evidence of the quenching in the Milky Way reported in \cite{Haywood2018}. The quenching that we find could be linked to a previous merger event. Simulations in the framework of $\Lambda CDM$ show that after a merger, there is an enhancement of the star formation followed by a quenching (e.g. \citealt{Dimate02008}, Fig. 4). This would be compatible with the thick-disc formation scenario recently proposed in \cite{Helmi2018} with a merger which occurred more than 10 Gyr ago. As suggested in other works, the quenching that we find could also be partially produced by the presence of a galactic bar \citep{HaywoodMateo2016,Khoperskov2018}. The two quenching mechanisms discussed here are not mutually exclusive but complementary. After the quenching, we detect a 4 Gyr duration SFR-enhancement event starting at about 5 Gyr ago and with a maximum at 2-3 Gyr ago. The large timescale of this recent SFR enhancement event, together with the large amount of mass that we estimate to be involved in it (see Sect. \ref{Sec5}), lead us to propose that this recent event is not intrinsic to the disc but is produced by an external perturbation. Furthermore, the slow increase of the star formation process, its duration, as well as the high absolute value of the maximum suggest that this could be produced by a recent merger with a gas-rich satellite galaxy that could have started between about 5 and 7 Gyr ago. However, an analysis of other stellar parameters (e.g. metallicities) would be needed to favour this hypothesis over other possible scenarios.

Work is in progress to more thoroughly analyse the Gaia DR2 data by extending our study to fainter magnitudes, updating the stellar evolutionary models and the thick disc modelling.

\begin{acknowledgements}
We thank the anonymous referee for the constructive report which helped to improve the quality of the work. Special thanks are to N. Lagarde, C. Charbonnel, and C. Reyl\'e for very useful discussions. This work was supported by the MINECO (Spanish Ministry of Economy) - FEDER through grant ESP2014-55996-C2-1-R and MDM-2014-0369 of ICCUB (Unidad de Excelencia 'Mar\'ia de Maeztu'),  the French Agence Nationale de la Recherche under contract  ANR-2010-BLAN-0508-01OTP and the  European Community's Seventh Framework Programme (FP7/2007-2013) under grant agreement GENIUS FP7 - 606740. We also acknowledge the team of engineers (GaiaUB-ICCUB) in charge to set up the Big Data platform (GDAF) at University of Barcelona. We also acknowledge the International Space Science Institute, Bern, Switzerland for providing financial support and meeting facilities.  This work has made use of data from the European Space Agency (ESA) mission {\it Gaia} (\url{https://www.cosmos.esa.int/gaia}), processed by
the {\it Gaia} Data Processing and Analysis Consortium (DPAC,
\url{https://www.cosmos.esa.int/web/gaia/dpac/consortium}). Funding
for the DPAC has been provided by national institutions, in particular
the institutions participating in the {\it Gaia} Multilateral Agreement.

\end{acknowledgements}

\bibliographystyle{aa}
\bibliography{ref2}

\begin{appendix}

\section{Treatment of the IMFs from \cite{Dib2018} and \cite{Jerabkova2018}}\label{AppIMF}

To be able to compare the IMFs of \cite{Dib2018} and \cite{Jerabkova2018} with the IMF that we obtain in this work we need to perform an adequate treatment. For the case of \cite{Dib2018}, to be able to compare the slopes, we fit a three truncated power-law IMF to their results when they assume $a_\Gamma=0.5$, $a_\gamma$=0.5, and $a_{M_{ch}}=0.5$ (see their Fig. 1).  We plot this result in Fig. \ref{IMF}. The case of \cite{Jerabkova2018} is slightly more complex as their IMF depends on the SFR and metallicity. From our results we estimate a mean SFR (in $M_\odot/yr$) and a mean metallicity for each one of the age bins considered. Then from \cite{Yan2017} and \cite{Jerabkova2018} we get a resulting IMF for each age bin (using galIMF; \url{https://github.com/Azeret/galIMF}). Finally we compute a weighted mean of the IMF depending on the total mass for each age bin. The total integrated galactic IMF is then plotted in Fig. \ref{IMF}.


\end{appendix}

\end{document}